\documentclass{ws-procs9x6-cpt22}
\begin{document}

\newcommand{\refeq}[1]{(\ref{#1})}
\def\etal {{\it et al.}}

\title{Lorentz Violation in Neutrino Oscillations using IceCube Atmospheric Neutrino Interferometry}

\author{B.\ Skrzypek$^{1}$ and C.\ A. Arg\"uelles,$^{1}$}

\address{$^1$Physics Department, Harvard University,\\
Cambridge, Massachusetts 02138, USA}



\begin{abstract}
Lorentz invariance is a fundamental symmetry of spacetime underpinning the Standard Model (SM) and our understanding of high-energy phenomena in particle physics. However, beyond the quantum gravity scale, we expect the SM to be replaced with a more fundamental, covariant theory giving a quantum description of gravity. The effective theory arising from this theory can break Lorentz invariance and thus predicts observables that exhibit low-energy manifestations of Lorentz violation. In particular, these observables could be a subleading contribution to neutrino oscillations and could therefore explain anomalous flavor measurements. The Standard Model Extension (SME) formalism describing such an effective theory predicts terms whose characteristic oscillation length becomes significant at atmospheric neutrino energies accessible by the IceCube Neutrino Observatory. We descibe past measurements and efforts to extend these using ten years of data along with a new energy reconstruction to study $\nu_{\mu}$ disappearance. 
\end{abstract}

\bodymatter

\section{Introduction}

Neutrino flavor oscillations are one of the first indications of physics beyond the Standard Model (SM), and to first order, they are explained by the $\nu$SM, which relates the neutrino flavor basis to a mass basis through a unitary transformation. 
Although the mechanism by which neutrinos obtain their mass is still largely unknown, the $\nu$SM can be formally written as
\begin{equation}
    \mathcal{L}_{\nu SM} = \frac{1}{2}(m_{\nu})_{ij}\nu_i^T\nu_j,
\end{equation}
where $m_{\nu}$ can be diagonalized to produce three neutrino mass eigenstates: 
\begin{equation}
    V_{\nu}^*m_{\nu}V_{\nu}^{\dagger} = {\rm diag}(m_1,m_2,m_3).
\end{equation}
This gives rise to the PMNS mixing matrix, which is the unitary transformation which relates the neutrino flavor basis to the neutrino mass basis: 
\begin{equation}
    U = V_{eL}V_{\nu L}^{\dagger}, \quad |\nu_{\alpha}\rangle = U_{\alpha i}^* |\nu_i\rangle.
\end{equation}
The $\nu$SM Lagrangian suggests that the neutrino mass scale, $\Lambda$ is much larger than the weak scale. Moreover, current experimental bounds on neutrino masses suggest an upper bound on the mass scale: 

\begin{equation}
    m_{1,2,3} \lesssim \frac{\nu^2}{\Lambda} \ll \nu \Rightarrow \Lambda \lesssim \frac{\nu^2}{m_{\nu}}\sim 10^{15} \, {\rm GeV},
\end{equation}
where here, $\nu$ is the weak Yukawa scale $\nu = (\sqrt{2}G_F)^{-1/2} \sim 246 \; \rm GeV.$
This suggests that the $\nu$SM cannot be valid beyond this scale, which motivates theories that alter neutrino oscillations at higher energies. 
The Standard Model Extension (SME) is one such framework that models Lorentz- and CPT-violation in the form of an expansion of operators that arise from potential spontaneous Lorentz-symmetry breaking of a covariant fundamental theory whose low-energy limit is the minimal SM \cite{Colladay:1996iz}. Our focus is on the lepton sector of the SME, which consists of CPT-odd and CPT-even terms: 

\begin{eqnarray}
    & \mathcal{L}_{\rm lepton}^{\rm CPT-even} = \frac{1}{2}i\,(\, c_L \,)_{\mu\nu AB}\,\bar{L}_A\,\gamma^{\mu}\,\overleftrightarrow{D}^{\nu}\,L_B + \frac{1}{2}i\,(\, c_R\, )_{\mu\nu AB}\,\bar{R}_A\,\gamma^{\mu}\,\overleftrightarrow{D}^{\nu}\,R_B, \nonumber \\
    & \mathcal{L}_{\rm lepton}^{\rm CPT-odd} = -(\, a_L\, )_{\mu AB} \,\bar{L}_A\,\gamma^{\mu}\,L_B - (\, a_R\, )_{\mu AB}\,\bar{R}_A\,\gamma^{\mu}\,R_B.
\end{eqnarray}

Neutrino experiments can constrain the SME by obtaining limits on the left-handed operators shown above. Without loss of generality, these can be assumed to be traceless, and we can expand the above Lagrangian to include higher-order terms as well. In the Hamiltonian picture, the interactions in the neutrino sector can be formally summarized in terms of the following generalized Hamiltonian, with effective operators appearing in each dimension \cite{Kostelecky:2011gq}: $H 
     =\frac{m^2}{2E}+\sum_{d\geq 3}p^{d-3}\big(a^{(d)}-c^{(d)})$.
Here, each coefficient represents a Lorentz-contraction (\textit{e.g.} $a^{(3)} \equiv a_{\mu}^{(3)}p^{\mu}$.)



\section{Neutrinos at IceCube}
The IceCube Neutrino Observatory is a 1 km$^3$ Cherenkov detector located in the South Pole. It consists of an array of 86 strings with 60 digital optical modules (DOMs) on each string. Each DOM is equipped with a photomultiplier tube that detects Cherenkov photons emitted by charged particles. There are generally three types of event morphologies which approximately correlate to the neutrino interaction type: \textit{Tracks} are indicative of charged-current $\nu_{\mu}$ events, \textit{cascades} originate from neutral current and charged-current $\nu_e$ events, and \textit{double cascades} result from charged-current $\nu_{\tau}$ events at around a PeV or higher. 

The majority of neutrinos that IceCube observes are atmospheric in nature, originating from cosmic-ray showers that interact with the Earth's atmosphere \cite{Vitagliano:2019yzm}.
This consists of neutrinos that come from pion, kaon, and charm meson decays. 
Because the cosmic-ray flux is approximately isotropic, atmospheric neutrinos traverse various distances through the earth before arriving at IceCube. 
As a result, tests of Lorentz violation in IceCube make use of the ratio between short- and long-baseline neutrino flux, as the latter is expected to exhibit a larger Lorentz-violating effect. 
\par Atmospheric neutrinos dominate the neutrino spectrum up to energies of around $60$ TeV, beyond which astrophysical neutrinos become the predominant component. Astrophysical neutrinos originate from processes taking place in high-energy environments such as star-forming galaxies, gamma-ray bursts, blazars, cosmogenic neutrinos, and beyond-SM processes such as dark-matter decay or annihilation. 
In order to avoid a large atmospheric background, astrophysical analyses typically rely on high energies and consider directions in which the Earth can block the atmospheric muons. 
However, the astrophysical neutrino spectrum has yet to be determined, as there are tensions among existing IceCube datasets \cite{IceCube:2020wum}. These tensions point to more a more complicated, multi-component astrophysical source. 

\section{Previous Searches using Atmospheric Neutrinos}

\begin{figure}
    \centering
    \includegraphics[width=0.4\textwidth]{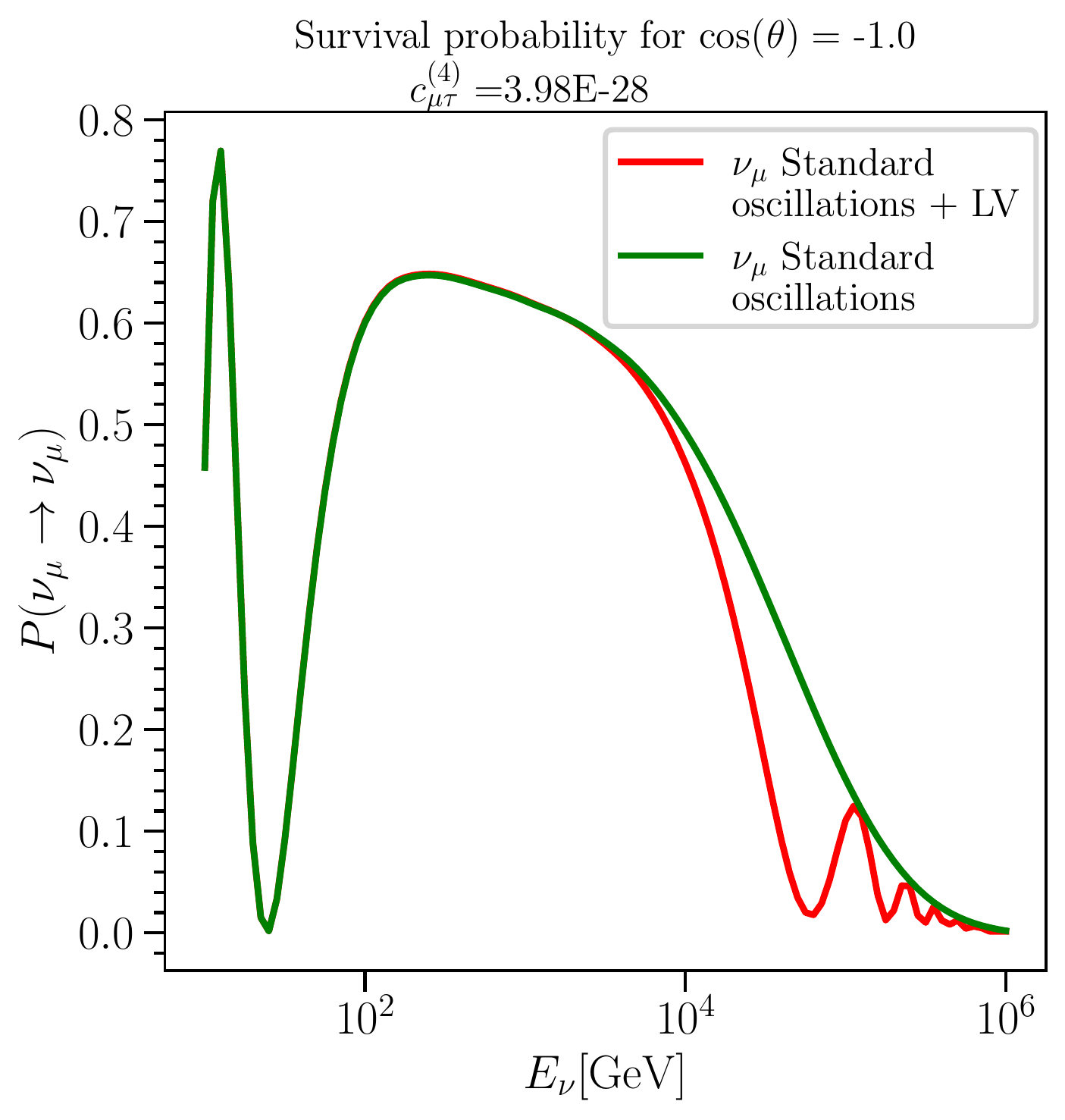}
    \includegraphics[width=0.45\textwidth]{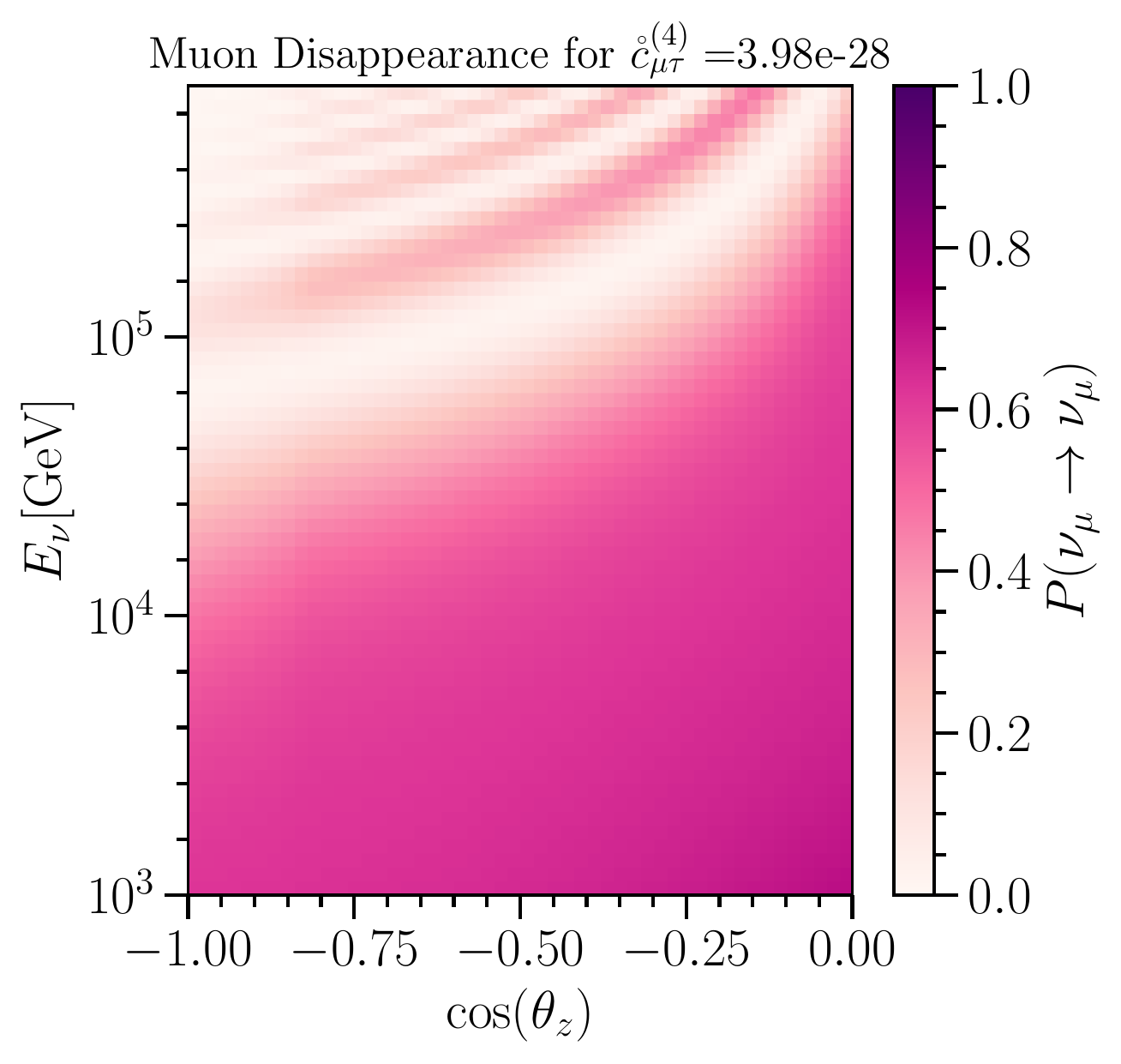}
    \caption{Left: The survival probability of a muon neutrino at IceCube as a function of its initial energy. Shown in green is the probability that results from standard oscillations, while the red curve is the deviation with Lorentz-violation. Here, we assume an SME scenario in which only the off-diagonal component of the dimension-4 coefficient is non-zero. We have included Earth-absorption into our calculation. Right: A two-dimensional oscillogram showing the survival probability of a muon neutrino as a function of its direction and energy. This is for the same SME scenario as shown on the left plot.}
    \label{fig:Oscillograms}
\end{figure}

A recent atmospheric search for Lorentz-violation used two years of data to obtain the strongest constraints across all experiments on the isotropic SME coefficients in the neutrino sector through dimension-8 \cite{IceCubeLV,Farrag:CPT22,IceCube:2021tdn}. In particular, the analysis yielded the most competitive limit for a dimension-6 operator across all sectors. This search used the earth as a neutrino interferometer, whereby neutrinos arriving from different directions will exhibit different oscillation patterns. This gives IceCube sensitivity to the characteristic oscillation length of a given SME coefficient, especially for high-dimensional operators, where the characteristic length scales inversely with energy. Under the assumption of maximal flavor violation, which is imposed by setting the diagonal terms of the SME coefficients to zero, the resulting transition probability is 
\begin{equation}
    P(\nu_{\mu}\rightarrow \nu_{\tau}) \sim \Big(\frac{a_{\mu\tau}^{(d)}-c_{\mu\tau}^{(d)}}{\rho_d}\Big)^2\sin^2\big(L\rho_d\cdot E^{d-3}\big),
\end{equation}
where $\rho_d = \sqrt{(c_{\mu\mu}^{(d)})^2+{\rm Re}(c_{\mu\tau}^{(d)})^2 + {\rm Im}(c_{\mu\tau}^{(d)})^2}$ is definition for the strength of the SME coefficient. 
The analysis was able to constrain the dimension-6 operator to $|c_{\mu\tau}|<1.5\times 10^{-36} \; {\rm GeV}^{-2}$ at $99\%$ confidence level. Figure~\ref{fig:Oscillograms} shows an example of a dimension-4 SME signature at IceCube, where we incorporate Earth absorption of muon neutrinos in our propagation. In contrast to standard oscillations, SME oscillations exhibit differences at high energies, particularly above 10 TeV.  

\section{New Search with Atmospheric \& Astrophysical Neutrinos}

In an attempt to improve these results further, we consider using the new eight years of IceCube data, which comes with an optimized event selection and improved systematic treatment. We also plan to investigate the astrophysical component, motivated by the recent quantum-gravity results from IceCube, in which the strongest limits on the dimension-6 SME operator were achieved to date. 

To improve our results, we use an optimized event selection with an enhanced DNN energy reconstruction and new variable cuts. When modeling the SME, we assume maximum-flavor violation and we only test for the isotropic component. 

In our likelihood analysis, we compute the likelihood differences between the null and SME hypotheses by using a Poisson likelihood. We then compute the sensitivity by assuming Wilk's theorem with one degree of freedom. Our preliminary results only include statistical errors and no systematic treatment, and we only assume a real SME coefficient for each dimension. For dimensions 4 and 5, the resulting statistics-only sensitivities show significant improvements from previous results. We can use these results to estimate that improving the energy cut from 16 TeV to 100 TeV would increase the sensitivity by a factor of about 5 (proportional to $E$) for dimension 4, and a factor of about 100 for dimension 6 (proportional to $E^3$). For more details about the sensitivities and the statistics-only results, please refer to the slides corresponding to this proceeding. 

In addition to the improved energy cuts, we anticipate the improvement in sensivity to remain when we incorporate detector systematics, an enhanced selection, and more data.
Moreover, in light of recent astrophysical results from IceCube that obtain the strongest constraints on the dimension-6 coefficient across all sectors to date \cite{quantumgravity}, we also plan to test the SME with astrophysical neutrinos. This introduces a new variable that describes the neutrino flavor composition at the astrophysical source. We plan to test whether the flavor composition of the resulting neutrinos that we observe lands outside of the region that is generally allowed by the Standard Model for astrophysical sources. This could allow us to further constrain certain SME parameters as a function of flavor-composition at the source. 



\section*{Acknowledgments}

CAA and BS are supported by the Faculty of Arts and Sciences of Harvard University.
Additionally, CAA thanks the Alfred P. Sloan Foundation for their support.

\end{document}